\documentclass[12pt,aps,prd,groupedaddress,preprint,tightenlines,floatfix,nofootinbib]{revtex4}
\usepackage{amssymb,amsmath,graphicx,multirow}

\newcommand{\PRE}[1]{{#1}} % Use if preprint style
\newcommand{\rb}[1]{\mathbf{#1}}
\newcommand{\rbb}[1]{\overline{\mathbf{#1}}}

\begin{document}

\preprint{UCI-TR-2011-17}
\preprint{UH-511-1176-11}

\title{ \PRE{\vspace*{0.8in}}
Higher Representations and Multi-Jet Resonances at the LHC
\PRE{\vspace*{0.3in}}
}
\author{Jason Kumar$^{1}$\footnote{E-mail address:  {\tt jkumar@hawaii.edu}},
      Arvind Rajaraman$^{2}$\footnote{E-mail address:  {\tt arajaram@uci.edu}},
      and Brooks Thomas$^{1}$\footnote{E-mail address:  {\tt thomasbd@phys.hawaii.edu}}}
\affiliation{
      $^1$ Department of Physics, University of Hawaii, Honolulu, HI 96822 USA\\
      $^2$ Department of Physics and Astronomy, University of California, Irvine, CA 92697 USA
\PRE{\vspace*{.5in}}
}

\begin{abstract}
 \PRE{\vspace*{.3in}}
  The CMS collaboration has recently conducted a search for trijet resonances
  in multi-jet events at the LHC.
  Motivated in part by this analysis, we examine the phenomenology of
  exotic particles transforming under higher representations of $SU(3)$ color,
  focusing on those representations which intrinsically prohibit decays to
  fewer than three jets.  We determine the LHC discovery reach for a particle
  transforming in a representation of this sort and discuss several additional
  theoretical and phenomenological constraints which apply to such a particle.
  Furthermore, we demonstrate that such a particle can provide a consistent
  explanation for a trijet excess (an invariant-mass peak of roughly
  375~GeV) observed in the recent CMS study.
\end{abstract}


\maketitle

%========================================================================
%          KEYSROKE-SAVING MACROS, nothing complicated
%========================================================================

\newcommand{\newc}{\newcommand}
\newc{\gsim}{\lower.7ex\hbox{$\;\stackrel{\textstyle>}{\sim}\;$}}
\newc{\lsim}{\lower.7ex\hbox{$\;\stackrel{\textstyle<}{\sim}\;$}}

\def\vac#1{{\bf \{{#1}\}}}

\def\beq{\begin{equation}}
\def\eeq{\end{equation}}
\def\beqn{\begin{eqnarray}}
\def\eeqn{\end{eqnarray}}
\def\calM{{\cal M}}
\def\calV{{\cal V}}
\def\calF{{\cal F}}
\def\half{{\textstyle{1\over 2}}}
\def\quarter{{\textstyle{1\over 4}}}
\def\ie{{\it i.e.}\/}
\def\eg{{\it e.g.}\/}
\def\etc{{\it etc}.\/}
%\def\alpha\dot{\alpha}{\alpha\dot{\alpha}}

%     The following macros are to create the "blackboard bold"
%     characters for "R" (set of real numbers),
%     "C" (set of complex numbers), and "Q" (set of rational numbers).

\def\inbar{\,\vrule height1.5ex width.4pt depth0pt}
\def\IR{\relax{\rm I\kern-.18em R}}
 \font\cmss=cmss10 \font\cmsss=cmss10 at 7pt
\def\IQ{\relax{\rm I\kern-.18em Q}}
\def\IZ{\relax\ifmmode\mathchoice
 {\hbox{\cmss Z\kern-.4em Z}}{\hbox{\cmss Z\kern-.4em Z}}
 {\lower.9pt\hbox{\cmsss Z\kern-.4em Z}}
 {\lower1.2pt\hbox{\cmsss Z\kern-.4em Z}}\else{\cmss Z\kern-.4em Z}\fi}
\def\wtg{\mathbf{g}}
\def\wtq{\mathbf{q}}
\def\st1{\mathbf{t}_1}
\def\mst1{m_{\mathbf{t}_1}}
\def\Neut1{\mathbf{N}_1}
\def\N1{\mathbf{N}_1}
\def\qbar{\overline{q}}
\def\etc{{\it etc}.\/}

\newcommand{\Dsle}[1]{\hskip 0.09 cm \slash\hskip -0.28 cm #1}
\newcommand{\met}{{\Dsle E_T}}
\newcommand{\mht}{{\Dsle H_T}}
\newcommand{\Dslp}[1]{\slash\hskip -0.23 cm #1}
\newcommand{\mpt}{{\Dslp p_T}}
\newcommand{\bigDsle}[1]{\hskip 0.05 cm \slash\hskip -0.38 cm #1}
\newcommand{\bigmet}{{\bigDsle E_T}}

\newcommand{\gev}{{\rm GeV}}
\newcommand{\tev}{{\rm TeV}}
\newcommand{\ifb}{{\rm fb^{-1}}}
\newcommand{\pb}{{\rm pb}}
\newcommand{\fb}{{\rm fb}}

%========================================================================

\input epsf

%========================================================================
%========================================================================
%               MAIN TEXT BEGINS HERE
%========================================================================

%========================================================================

%-----------------------------------------------------------------------------
\section{Introduction}
%-----------------------------------------------------------------------------

The advent of data from the Large Hadron Collider (LHC) has already begun
to provide a meaningful probe into a wide variety of
long-standing scenarios for new physics.
Even with only $\mathcal{L}_{\mathrm{int}}\approx 1\mathrm{~fb}^{-1}$
of data currently under analysis, the ATLAS and CMS experiments have been able to
place stringent constraints on many extensions of the Standard Model (SM).  At such integrated
luminosities, the processes to which these experiments are sensitive are most
notably those in which new particles are produced either via strong interactions,
or else through an $s$-channel resonance.  Indeed, from limits on processes of this
sort, LHC data have already placed stringent constraints on the parameter
space of many of the most widely studied extensions of the SM,
including many models involving weak-scale supersymmetry, extra dimensions,
and additional exotic states such as $Z'$ gauge bosons.

In addition to these popular scenarios, it is worthwhile to
look for signals of less traditional extensions of the SM
which, for one reason or another, could have been missed by the standard
battery of new-physics searches at the LHC.  For example, scenarios exist in which a
strongly-interacting particle is produced copiously in hadron collisions, but
decays in unexpected ways and is consequently overlooked.
One example of such a particle is a light gluino in a supersymmetric
theory with R-parity violation.  Such a particle has a relatively large
pair-production cross-section at the LHC; however, since each gluino
so produced decays predominately to three jets, evidence for an
R-parity-violating gluino would appear only in events with six or more
jets in the final state.
Typical searches do not consider such high jet multiplicities, and based upon the
results of those searches alone, it is almost inevitable that any particle with
a decay pattern of this sort would be overlooked.
However, searches for multi-jet
resonances in high-jet-multiplicity events could potentially reveal evidence
of such a particle.  Motivated by this consideration,
the CMS experiment recently performed a study of the three-jet invariant
mass distribution in events with at least six jets~\cite{CMSTrijetSearch} with
$35.1~\mathrm{ pb}^{-1}$ of LHC data.
The results of this study (about which we
will say more later) now provide the leading constraints on the
gluino mass in models with R-parity violation.  A similar analysis was also
recently performed by the CDF collaboration~\cite{CDFTrijetSearch} with
$3.2~\mathrm{ fb}^{-1}$ of Tevatron data.

Information about supersymmetry is not the only aspect of
physics beyond the standard model
into which searches for resonances in multi-jet processes could
provide an important window.  For example, one of the most
fundamental questions in particle physics is whether the SM
gauge interactions unify at some high scale, and if so, precisely how
this unification takes place and at what scale that might be.  Experimental
signals which could provide information on unification are therefore
immensely valuable from a theoretical perspective.  An example of such a
signal would be the discovery of an exotic matter field charged under the
SM $SU(3)_c$ gauge group.  Indeed, a particle of this sort would significantly
alter the renormalization-group running of the strong coupling coefficient
$\alpha_s$, particularly if the $SU(3)_c$ representation under which those
matter fields transformed was one of particularly large dimension.  Thus, the
presence of such a field would compel a revision of our projection for the scale of
grand unification --- potentially dramatically, if the dimension
of the representation in which the field or fields transformed were large enough to
spoil the asymptotic freedom of $SU(3)_c$.
Furthermore, as discussed above, strongly-interacting particles of this sort are
precisely the sort of new physics to which LHC data will be sensitive during the
first few $\mathrm{fb}^{-1}$ of running.

The prospects for observing exotic particles transforming in certain
higher representations of $SU(3)_c$ at hadron colliders in final states
comprising either four or six jets have been discussed in the
literature before~\cite{SehkarMultijet,EssigThesis}.
Moreover, a number of searches for new strongly-interacting particles, including
gluinos~\cite{CDFSUSYSearch,D0SUSYSearch,CMSSUSYSearch,ATLASSUSYSearch,ATLASSUSYSearchSimplifiedModel}, diquarks~\cite{ATLASDijetSearch,CMSDijetDiquarkSearch1,CMSDijetDiquarkSearch2},
fourth-generation quarks~\cite{D04thGenQuarks,CDF4thGenQuarks,ATLAS4thGenQuarks}, and miscellaneous
color-octet, sextet, and triplet
fields~\cite{ATLASDijetSearch,CMSDijetDiquarkSearch1,CMSDijetDiquarkSearch2}
have been performed both at the Tevatron and at the LHC.  To date, no
compelling evidence of such particles has been found.
However, such searches are generally only sensitive to the presence of
particles which can decay either to a pair
of strongly-interacting SM fields (quarks or gluons), or else to a final state
including some lighter neutral field which appears as missing energy.  By contrast,
a strongly-interacting particle which is forbidden by symmetry from coupling to any
pair of SM fields in a theory in which no lighter, neutral field exists will be
unconstrained by bounds from these typical searches.  Indeed, this is precisely the
case for the R-parity-violating gluino scenario discussed above.

However, a gluino of this sort is by no means the only example of a particle which might
have been overlooked in traditional searches for new strongly-interacting fields.
For example, as we shall demonstrate, there exist particular representations of $SU(3)_c$
for which an exotic field $X$, if it transforms under one of these representations, is
forbidden by gauge invariance from coupling to any pair of SM fields, but {\it can} couple
to at least one combination of three SM quarks or gluons.  The primary decay channel
for such a field would likewise therefore be to three jets, while a two-jet final state would
be forbidden.

We have already mentioned some of the theoretical motivations for examining the detection
prospects for fields which transform under higher representations of $SU(3)_c$.
In addition, there is now also a motivation for such an analysis from LHC data.  The
aforementioned CMS study~\cite{CMSTrijetSearch} did observe an excess in
the trijet-invariant-mass distribution at $M_{jjj} \sim 375$~GeV which differs
from the SM prediction by more than $2\sigma$ (though the significance is reduced to
$1.9\sigma$ once the look-elsewhere effect is included).  The authors compared this observed
excess to that which would result from the R-parity-violating decays of a gluino with a mass
$M_{\widetilde{g}}=375$~GeV, which turns out to be too small by a factor of roughly three.
In other words, for some other field to provide a more compelling explanation of
the observed excess, the product of pair-production cross-section for that field
would need to be approximately thrice the $\sim 15~\pb$ expected for a
gluino with $M_{\widetilde{g}}=375$~GeV, assuming the branching fraction for that field
into three jets is roughly unity.  Indeed, as we shall demonstrate, a particle $X$ transforming
in a higher representation of $SU(3)_c$ is capable of yielding an excess of the
observed magnitude.

We begin our analysis of the collider phenomenology of an exotic field
transforming under a higher representation of $SU(3)_c$ in
Sect.~\ref{sec:RepTheory} by examining the decay
properties of such a field from a representation-theory perspective.
In particular, we determine the representations for which the
transformation properties of such a field under $SU(3)_c$ and spacetime
symmetries alone forbid all direct decays to states involving only two SM quarks or
gluons, but permit at least one decay channel involving three such particles.  In
Sect.~\ref{sec:Pheno}, we investigate the collider phenomenology of exotic
fields in such representations, in which (again, due to considerations
related to representation theory) pair production via strong interactions plays
the dominant role.
In Sect.~\ref{sec:CMSComparison}, we discuss the implications of the recent
CMS multi-jet resonance search for fields in higher representations
of $SU(3)_c$ and compare our results for the signals expected from such fields
to the excess reported in Ref.~\cite{CMSTrijetSearch}.  In
Sect.~\ref{sec:Conclusions}, we conclude.

%-----------------------------------------------------------------------------
\section{Representation Theory and Decays to Three Jets\label{sec:RepTheory}}
%-----------------------------------------------------------------------------

Our primary aim in this paper is to examine the multi-jet phenomenology
of an exotic field $X$ transforming under a higher representation of $SU(3)_c$.
However, in order to do this,
we must first establish for which representations such a field
is forbidden from coupling directly to any pair of SM particles by $SU(3)_c$ gauge
invariance and Lorentz invariance alone, but for which at least one
gauge-invariant coupling to three strongly-interacting SM fields exists.
In other words, we wish to enumerate the $SU(3)_c$ representations which
permit at least one gauge-invariant operator $\mathcal{O}_i^{(3)}$ of the form
\begin{equation}
  \mathcal{O}_i^{(3)} ~ = ~ \frac{C_i^{(3)}}{\Lambda^{n_i}}
       X\tilde{\mathcal{O}}_i^{(3)}(g,q,\qbar)~,
  \label{eq:Ops3Body}
\end{equation}
where $\tilde{\mathcal{O}}_i^{(3)}(g,q,\overline{q})$ is an operator consisting
of exactly three SM fields charged under $SU(3)_c$
(\ie, quarks, antiquarks, or gluons), $C_i^{(3)}$ is a dimensionless operator coefficient,
$\Lambda$ is the suppression scale for the operator, and the value of the integer
$n_i$ depends on the whether $X$ is a scalar or a fermion and on the particular collection
of SM fields out of which $\tilde{\mathcal{O}}_i^{(3)}(g,q,\qbar)$ is constructed.  At
the same time, we require that there not exist any gauge-invariant operator of the form
\begin{equation}
  \mathcal{O}_j^{(2)} ~ = ~ \frac{C_j^{(2)}}{\Lambda^{n_j}}
       X\tilde{\mathcal{O}}_j^{(2)}(g,q,\qbar)~,
  \label{eq:Ops2Body}
\end{equation}
where $\tilde{\mathcal{O}}_j^{(2)}(g,q,\qbar)$ is an operator consisting
of exactly two SM fields charged under $SU(3)_c$, and $C_j^{(2)}$ is, once again,
a dimensionless coefficient.

In Table~\ref{tab:ParticleDecomp}, we list the $SU(3)_c$ representations which
appear at least once in the decomposition of each possible direct product of two or
three factors, assuming each factor is a $\rb{3}$, $\rbb{3}$, or $\rb{8}$
representation of $SU(3)_c$.  In addition, we display the Lorentz representations
which can be built from each of the corresponding three-particle states,
up to spin 1.  We see from this table that a number of representations
exist which do not appear in the decomposition of any two-particle state,
but exist in the decomposition of at least one three-particle state.  These
include the complex representations $\rb{15}'$, $\rb{24}$, $\rb{35}$, and
$\rb{42}$ (and their conjugate representations), as well as the real
representation $\rb{64}$.\footnote{Note that two distinct fifteen-dimensional
representations of $SU(3)_c$ exist, which we refer to here as $\rb{15}$ and
$\rb{15}'$.  The latter of these designates the completely symmetric
combination of four $\rb{3}$ representations.}
Moreover, we see that if $X$ transforms in the $\rb{10}$ of $SU(3)_c$, it can
decay to two SM fields only if it is a boson.  A fermionic $\rb{10}$
must therefore decay to at least three strongly-interacting SM fields and,
by similar reasoning, so must a bosonic $\rb{15}$.

\begin{table}[ht!]
\begin{center}
\begin{tabular}{|c|c|cc|}\hline
   \multirow{2}{*}{~Final state~}&\multirow{2}{*}{~Product~}&
   \multicolumn{2}{|c|}{~~~~~Representations}\\
   & &~$SU(3)_c$~&~~~~~Lorentz~~~~~\\ \hline
   $q\qbar     $&$\rb{3} \otimes \rbb{3} $&
               $\rb{1},\rb{8}                                                        $~&S,V\\
   $qq         $&$\rb{3} \otimes \rb{3}  $&
               $\rbb{3},\rb{6}                                                       $~&S,V\\
   $\qbar\qbar $&$\rbb{3} \otimes \rbb{3}$&
               $\rb{3},\rbb{6}                                                       $~&S,V\\
   $qg         $&$\rb{3} \otimes \rb{8}  $&
               $\rb{3},\rbb{6},\rb{15}                                               $~& F \\
   $\qbar g    $&$\rbb{3} \otimes \rb{8} $&
               $\rbb{3},\rb{6},\rbb{15}                                               $~& F \\
   $gg         $&$\rb{8} \otimes \rb{8}  $&
               $\rb{1},\rb{8},\rb{10},\rbb{10},\rb{27}                               $~&S,V\\
   \hline
   $qqq            $~&$\rb{3} \otimes \rb{3} \otimes \rb{3}   $~&
                   $~\rb{1},\rb{8},\rb{10}                                           $~& F \\
   $qq\qbar        $~&$\rb{3} \otimes \rb{3} \otimes \rbb{3}  $~&
                   $~\rb{3},\rbb{6},\rb{15}                                          $~& F \\
   $q\qbar\qbar    $~&$\rb{3} \otimes \rbb{3} \otimes \rbb{3} $~&
                   $~\rbb{3},\rb{6},\rbb{15}                                         $~& F \\
   $\qbar\qbar\qbar$~&$\rbb{3} \otimes \rbb{3} \otimes \rbb{3}$~&
                   $~\rb{1},\rb{8},\rbb{10}                                          $~& F \\
   $gqq            $~&$\rb{8} \otimes \rb{3} \otimes \rb{3}  $~&
                   $~\rbb{3},\rb{6},\rbb{15},\rb{24}                                 $~&S,V\\
   $gq\qbar        $~&$\rb{8} \otimes \rb{3} \otimes \rbb{3}  $~&
                   $~\rb{1},\rb{8},\rb{10},\rbb{10},\rb{27}                          $~&S,V\\
   $g\qbar\qbar    $~&$\rb{8} \otimes \rbb{3} \otimes \rbb{3} $~&
                   $~\rb{3},\rbb{6},\rb{15},\rbb{24}                                 $~&S,V\\
   $ggq            $~&$\rb{8} \otimes \rb{8} \otimes \rb{3}   $~&
                   $~\rb{3},\rbb{6},\rb{15},\rb{15}',\rbb{24},\rb{42}                $~& F \\
   $gg\qbar        $~&$\rb{8} \otimes \rb{8} \otimes \rbb{3}  $~&
                   $~\rbb{3},\rb{6},\rbb{15},\rbb{15}',\rb{24},\rbb{42}              $~& F \\
   $ggg            $~&$\rb{8} \otimes \rb{8} \otimes \rb{8}   $~&
                   $~\rb{1},\rb{8},\rb{10},\rbb{10},\rb{27},\rb{35},\rbb{35},\rb{64} $~&S,V\\
   \hline
\end{tabular}
\caption{A list of all representations contained in the
decomposition of all combinations of two or three of the
particles $q$, $\overline{q}$, and $g$.
Repeated occurrences of any single representation in the
decomposition of each such product have been suppressed.
\label{tab:ParticleDecomp}}
\end{center}
\end{table}

In Table~\ref{tab:RepsAndCasimirs}, we list all the
combinations of $SU(3)_c$ and Lorentz representations for a field $X$
which prohibit its decays to all final states comprising only two quarks
or gluons, but permit decays to at least one final state comprising three such
particles.  Note that we do not
impose any additional restriction on the coupling structure of $X$ based on
its $U(1)_{\mathrm{EM}}$ charge $Q_X$; rather, we require that all operators of the
form given in Eq.~(\ref{eq:Ops2Body}) be excluded on the basis of
$SU(3)_c$ and Lorentz structure alone, and then assign $X$ whatever
electromagnetic charge is required by gauge invariance.  We find that
the smallest-dimension $SU(3)_c$ representation for $X$ for which this
condition is satisfied (aside from of course a fermionic octet, for which
the gluino is the prototypical example) is the $\rb{10}$, for which $X$ must
be fermionic (or otherwise it could decay to a pair of gluons).  We also find
that no representations with dimension greater than 64 can decay to three SM
fields alone.

\begin{table}[ht!]
\begin{center}
\begin{tabular}{|cc|c|cc|}\hline
   \multicolumn{2}{|c|}{~~~~~~~Representation~~~~~~~} & \multirow{2}{*}{~$Q_X$~} &
   \multirow{2}{*}{~$C_2(r)$~}&\multirow{2}{*}{~$C(r)$~}\\
   ~~~~~$SU(3)_c$~ &~Lorentz~& & & \\ \hline
  ~~~~$\mathbf{10},\mathbf{\overline{10}}$~  & F
               &~$+2,+1,0,-1    $~ & 6    & 15/2~  \\
  ~~~~$\mathbf{15},\mathbf{\overline{15}}$~  & S,V
               &~$+\frac{2}{3},-\frac{1}{3},-\frac{4}{3}$~ & 16/3 & 10~    \\
  ~~~~$\mathbf{15}',\mathbf{\overline{15}}'$~& F
               &~$+\frac{2}{3},-\frac{1}{3}     $~ & 28/3 & 35/2~  \\
  ~~~~$\mathbf{24},\mathbf{\overline{24}}$~  & S,F,V
               &~$+\frac{4}{3},+\frac{1}{3},-\frac{2}{3}$~ & 25/3 & 25~    \\
  ~~~~$\mathbf{35},\mathbf{\overline{35}}$~  & S,V
               &~$0             $~ & 12   & 105/2~ \\
  ~~~~$\mathbf{42},\mathbf{\overline{42}}$~  & F
               &~$+\frac{2}{3},-\frac{1}{3}     $~ & 34/3 & 119/2~ \\
  ~~~~$\mathbf{64}$~                         & S,V
               &~$0             $~ & 15   & 120~   \\ \hline
\end{tabular}
\caption{A list of combined $SU(3)_c$ and Lorentz representations for a hypothetical
particle $X$ for which the effective couplings between $X$ and all two-particle
combinations of $g$, $q$, or $\overline{q}$ are forbidden by symmetries, while
an effective coupling between at least one three-particle combination of these same
fields is allowed.  The invariants $C_{2}(r)$ (\ie, the quadratic
Casimir) and $C(r)$ for these representations are also given.  In addition, a list of
$U(1)_{\mathrm{EM}}$ charges $Q_X$ for which at least one gauge-invariant
$\mathcal{O}^{(3)}_i$ can be constructed is provided for each choice of $SU(3)_c$ and
Lorentz representations shown.  It should be noted that for the $\rb{24}$, the specific
assignment $Q_X = +\frac{4}{3}$ is consistent for a scalar or vector,
but not for a fermion.
\label{tab:RepsAndCasimirs}}
\end{center}
\end{table}

\begin{figure}[t!]
\centerline{
  \epsfxsize 3.25 truein \epsfbox {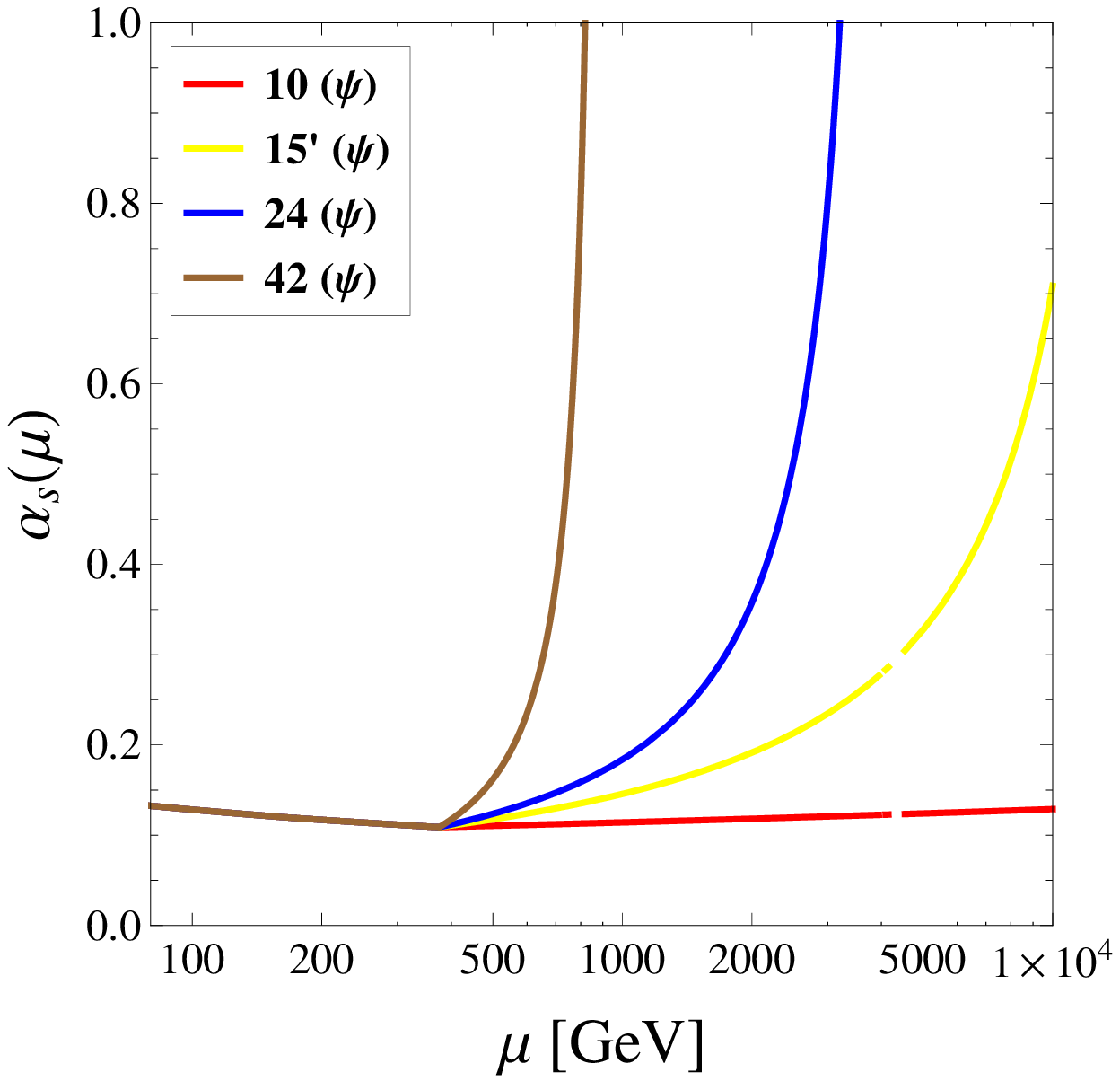}~~~~
  \epsfxsize 3.25 truein \epsfbox {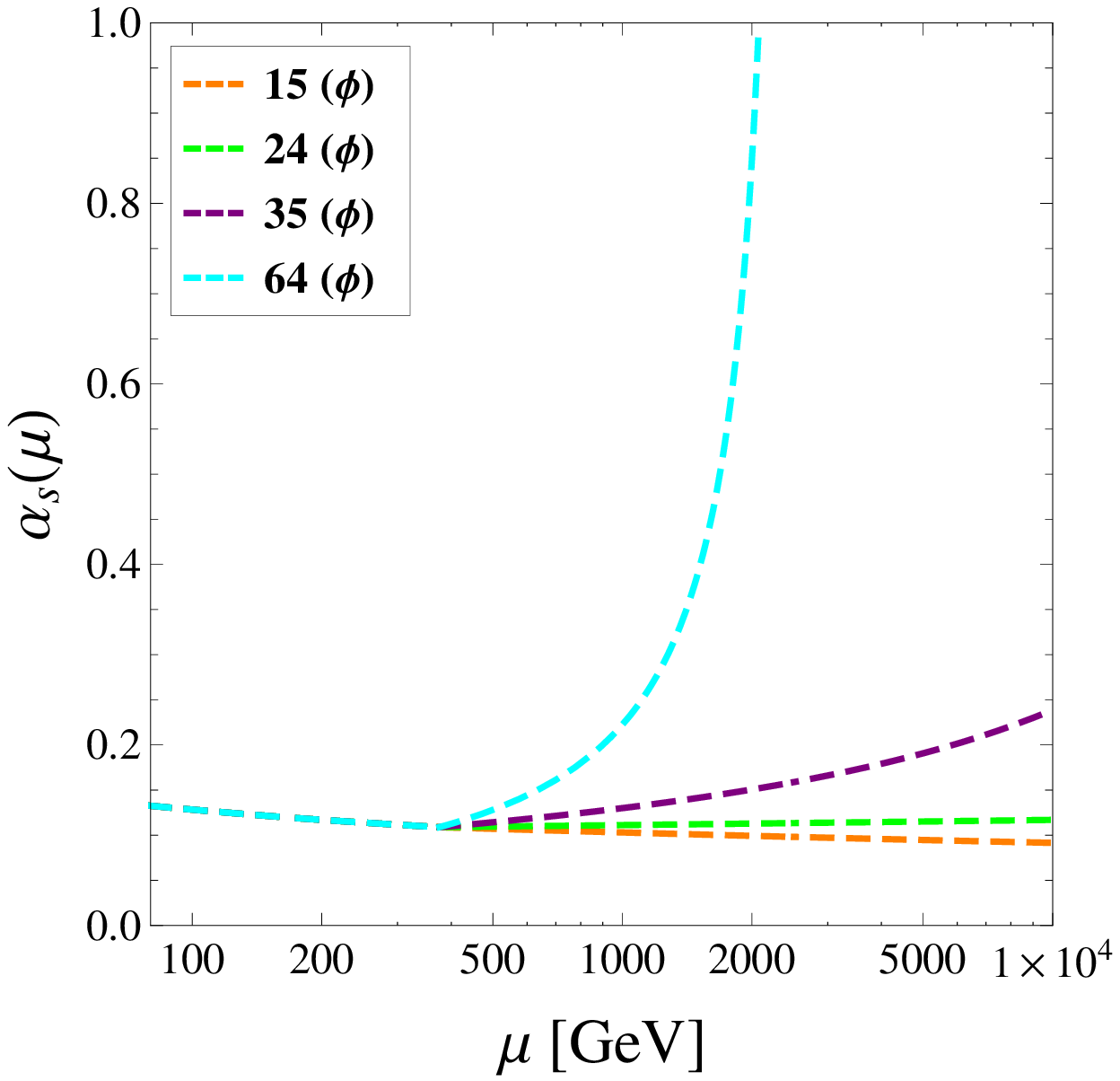} }
\caption{Curves indicating the renormalization-group evolution of $\alpha_s$
in the presence of a single exotic field $X$ in each of the representations
of $SU(3)_c$ enumerated in Table~\ref{tab:RepsAndCasimirs}.  The results
in the left panel correspond to the cases in which $X$ is a Dirac fermion, while
the results in the right panel correspond to the cases in which $X$ is a
scalar.  In each case, we have taken $m_X = 375$~GeV.
\label{fig:AlphasPlots}}
\end{figure}

One constraint on the presence of additional fields charged under $SU(3)_c$
comes from the requirement that $\alpha_s \equiv g_s^2/(4\pi)$ must be well-behaved
up to around the TeV scale.  In the presence of an exotic field $X$ charged under
an arbitrary representation of $SU(3)_c$, $\alpha_s(\mu)$ is modified at scales
$\mu > m_X$ to
\begin{equation}
  \alpha_s(\mu) ~=~ \frac{\alpha_s(M_Z)}{\displaystyle 1+\frac{\alpha_s(M_Z)}{12\pi}
  \Bigg[23\ln\left(\frac{\mu^2}{M_Z^2}\right)
     -2\ln\left(\frac{\mu^2}{m_t^2}\right)
  -\frac{f_X}{8}C(r)d(r)\ln\left(\frac{\mu^2}{m_X^2}\right)\Bigg]}~,
\end{equation}
where $\alpha_s(M_Z)\approx 0.118$ is the value of $\alpha_s(\mu)$ at
$\mu = M_Z \approx 91.19$~GeV
and $f_X = 1$ ($f_X = 4$) when $X$ is a scalar field (Dirac fermion).
The effect of including an additional field $X$ which transforms under each of the
representations of $SU(3)_c$ listed in Table~\ref{tab:RepsAndCasimirs} is shown in
Fig.~\ref{fig:AlphasPlots}.  The curves shown in the left panel correspond to the
cases in which $X$ is a scalar, while the curves in the right panel correspond to the
cases in which $X$ is a  Dirac fermion.  For each case, we have assumed that only a
single additional generation of $X$ is present, and we have taken $m_X = 375$~GeV.

It is evident from Fig.~\ref{fig:AlphasPlots} that the presence of even a single
field $X$ transforming in many of the representations listed in
Table~\ref{tab:RepsAndCasimirs} will result in $\alpha_s(\mu)$ developing
a Landau pole at a scale
$\mu \sim \mathcal{O}(\mathrm{TeV})$.  Imposing for theoretical consistency
the requirement that such a divergence not appear at scales below $\mu \sim 5$~TeV,
we find that both a fermionic $\rb{24}$ and $\rb{42}$, as well
as a scalar $\rb{64}$, are excluded.   However,
this constraint does not exclude a fermionic $\rb{10}$ or $\rb{15}'$, nor does it
exclude a scalar $\rb{15}$, $\rb{24}$, or $\rb{35}$.
For the rest of the paper we will therefore focus exclusively on these latter
representations of low dimension.  We note that
for a single  scalar $\rb{15}$,  the theory remains asymptotically free, but
for the other representations,
a Landau pole develops for $\alpha_s(\mu)$
at some scale $\mu > 10$~TeV; we assume that
a suitable short distance theory regulates this divergence.

Two additional comments are in order concerning the case in which $X$ is a
fermion.  First, the permissible $SU(3)_c$ representations
for a fermionic $X$, namely $\rb{10}$ and $\rb{15}'$, are both complex; this
implies that a fermion transforming in any of these representations must be a
Dirac rather than a Majorana particle.  Second,  whenever $X$ has
chiral charges, anomaly-cancellation requirements place additional constraints
on the theory.  Given the unorthodox representation of $SU(3)_c$ in which $X$ is
assumed to transform, these constraints are generally quite difficult to satisfy
simultaneously in any phenomenologically reasonable model.  We will
henceforth assume that $X$ is vector-like and thus does not contribute to gauge
anomalies.

%-----------------------------------------------------------------------------
\section{Collider Phenomenology\label{sec:Pheno}}
%-----------------------------------------------------------------------------

Having now established the representations of $SU(3)_c$ for which an exotic
field can decay to no fewer than three jets, we proceed to investigate
the collider phenomenology of a field $X$ transforming in one of these
representations.  We focus here on the case in which $X$ is either a scalar or
a Dirac fermion.  Since, by construction, no gauge-invariant operators of the form
specified in Eq.~(\ref{eq:Ops2Body}) exist for $X$, no
couplings of the form $ggX$,
$qqX$, or $\overline{q}qX$ exist either.  It therefore follows that $X$
cannot be produced singly as an $s$-channel resonance.
It may be produced in association with some other SM particle or
particles through an operator of the form given in Eq.~(\ref{eq:Ops3Body}),
but the corresponding
amplitude would be suppressed by powers of $\Lambda$.
As a result,
the pair production of $X$ and $\overline{X}$ (via the coupling
$g\overline{X}X$ to the gluon field required by gauge invariance)
is the dominant production channel at hadron colliders.

Once produced, we assume that
$X$ decays exclusively via operators of the form $\mathcal{O}_i^{(3)}$
to a trijet final state, with $\mathrm{BR}(X\rightarrow jjj)\approx 1$.  Indeed, by
construction, all two-body decay channels for $X$ are forbidden.  Moreover, restrictions
on the permissible $Q_X$ assignments for $X$ detailed in Table~\ref{tab:RepsAndCasimirs}
imply that all additional three-body decays involving charged leptons are forbidden by
charge conservation for all viable $SU(3)_c$ and Lorentz representations of $X$, save
for potentially the fermionic $\rb{10}$.  Even for this representation, such decays may
be forbidden either by choosing $Q_X = +2$ or by requiring lepton-number conservation and
assigning $X$ a lepton number $L_X = 0$.  Consequently, we expect the primary collider
signature of $X$ to be analogous to that of an R-parity-violating gluino:
a final state consisting of at least six high-$p_T$ jets, from which two combinations of
three jets reconstruct to an invariant mass peak at $M_{jjj}\approx m_X$.  Moreover,
since we are assuming that $\mathrm{BR}(X\rightarrow jjj)\approx 1$,
the collider phenomenology of $X$ will be essentially independent
of the operator coefficients $C_i^{(3)}$ and the suppression scale $\Lambda$ appearing
in Eq.~(\ref{eq:Ops3Body}) , as long as  these quantities are such that
$X$ decays promptly.

We begin our analysis of the process $pp\rightarrow\overline{X}X\rightarrow N_j$~jets,
where $N_j \geq 6$, at the LHC by deriving expressions for the production cross-section
of a scalar or
fermionic field $X$ in an arbitrary representation $r$ of $SU(3)_c$ with dimension $d(r)$.
For the case in which $X$ is a scalar, the leading-order (LO) partonic cross-sections
for the pair production of $X$ from the $q\overline{q}$ and $gg$ initial
states are
\begin{eqnarray}
  \hat{\sigma}_{q\overline{q}\rightarrow\overline{X}X}(\hat{s}) & = &
     \frac{\pi\alpha_s^2}{54\hat{s}}C_2(r) d(r) R^3\nonumber\\
 \hat{\sigma}_{gg\rightarrow\overline{X}X}(\hat{s}) & = &
   \frac{\pi\alpha_s}{64\hat{s}}C_2(r)d(r)\Bigg(
   \left[2\left(1+\frac{4m_X^2}{\hat{s}}\right)C_2(r)
   -1+\frac{10m_X^2}{\hat{s}}\right]R
   \nonumber \\ & & ~~~~~~~~~~~~~~~~~~~~~~
   -8\frac{m_X^2}{\hat{s}}\left[\frac{3m_X^2}{\hat{s}}
   +\left(1 - \frac{2m_X^2}{\hat{s}}\right)C_2(r)\right]
   \ln\left(\frac{1+R}{1-R}\right)\Bigg)~,
  \label{eq:ScalXProdXsecs}
\end{eqnarray}
where $C_2(r)$ is the quadratic Casimir associated with $r$,
$m_X$ is the mass of $X$, $\hat{s}$ is the partonic center-of-mass energy,
and $R \equiv \sqrt{1-4m_X^2/\hat{s}}$.
By contrast, for the case in which $X$ is a Dirac fermion, the corresponding
partonic cross-sections are found to be
\begin{eqnarray}
  \hat{\sigma}_{q\overline{q}\rightarrow\overline{X}X}(\hat{s}) & = &
     \frac{2\pi\alpha_s^2}{27\hat{s}}C_2(r)d(r)\left(1+\frac{2m_X^2}{\hat{s}}\right)R
     \nonumber\\
 \hat{\sigma}_{gg\rightarrow\overline{X}X}(\hat{s}) & = &
    \frac{\pi\alpha_s^2}{16\hat{s}}C_2(r)d(r)
    \Bigg(\left[\left(1+\frac{4m_X^2}{\hat{s}}-\frac{8m_X^4}{\hat{s}^2}\right)C_2(r) +
    \frac{12m_X^4}{\hat{s}^2}\right]\ln\left(\frac{1+R}{1-R}\right)
    \nonumber\\ & & ~~~~~~~~~~~~~~~~~~~~~~
    -\left[\left(1+\frac{4m_X^2}{\hat{s}}\right)C_2(r)+
    1+\frac{5m_X^2}{\hat{s}}\right]R\Bigg)~.
   \label{eq:FermXProdXsecs}
\end{eqnarray}
Note that since the gluino is a Majorana fermion, the partonic cross-sections for
gluino production~\cite{DawsonXSec,KuleszaXSec} are smaller than those obtained
from Eq.~(\ref{eq:FermXProdXsecs}) for an $\rb{8}$ by a factor of two.
The total LO production cross-section for the pair-production of
$X$ and $\overline{X}$ at the LHC can be written in the form
\begin{equation}
  \sigma_{pp\rightarrow \overline{X}X}(s) ~=~ \sigma_{q\bar{q}\rightarrow \overline{X}X}(s)
     + \sigma_{gg\rightarrow \overline{X}X}(s)~,
   \label{eq:TotXSec}
\end{equation}
where $\sigma_{q\bar{q}\rightarrow \overline{X}X}(s)$ and
$\sigma_{gg\rightarrow \overline{X}X}(s)$ denote the results of convolving
the partonic cross-sections in Eqs.~(\ref{eq:ScalXProdXsecs})
and~(\ref{eq:FermXProdXsecs}) with the appropriate parton-distribution
functions (PDFs) $f_{p/q}(x,Q^2)$, $f_{p/\bar{q}}(x,Q^2)$, and $f_{p/g}(x,Q^2)$:
\begin{eqnarray}
  \sigma_{q\bar{q}\rightarrow \overline{X}X}(s) &\equiv&2\sum_{q=u,d,s,c}
    \int_{\tau_0}^s\int_{\tau}^1 \frac{1}{\tau}
    f_{p/q}(x,\tau s)f_{p/\bar{q}}(\tau/x,\tau s)
    \hat{\sigma}_{q\bar{q}\rightarrow \overline{X}X}(\tau s) dx d\tau
    \nonumber\\
  \sigma_{gg\rightarrow \overline{X}X}(s) &\equiv&
    \int_{\tau_0}^s\int_{\tau}^1 \frac{1}{\tau}f_{p/g}(x,\hat{s})f_{p/g}(\tau/x,\hat{s})
    \hat{\sigma}_{q\bar{q}\rightarrow \overline{X}X}(\tau s) dx d\tau~.
    \label{eq:ggandqqtoXXxsectot}
\end{eqnarray}

\begin{figure}[ht!]
\centerline{
  \epsfxsize 3.25 truein \epsfbox {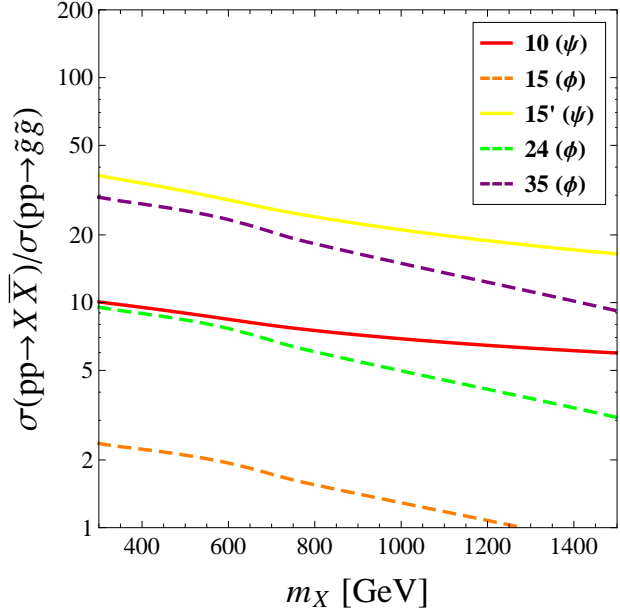} }
\caption{Ratios of $\sigma(pp\rightarrow X\overline{X})$ to
the gluino-pair-production cross section
$\sigma(pp\rightarrow\widetilde{g}\widetilde{g})$ at leading-order and in the
limit where all squark masses $m_{\overline{q}}$ are taken to infinity.
The solid lines correspond to the cases in which $X$ is a (Dirac)
fermion; the dashed lines correspond to the cases in which $X$ is a
scalar.\label{fig:EnhVsGluinoPlot}}
\end{figure}

In Fig.~\ref{fig:EnhVsGluinoPlot}, we plot the ratio of
$\sigma_{pp\rightarrow \overline{X}X}(s)$ to the total production cross-section
for a gluino of the same mass at leading-order (in the limit in which all squark masses are
taken to be infinitely heavy) as a function of $m_X$.  The curves shown
correspond to all otherwise phenomenologically consistent combinations of
$SU(3)_c$ and Lorentz representations for $X$ for which all
$\mathcal{O}_i^{(2)}$ are forbidden, but for which at least one
$\mathcal{O}_i^{(3)}$ is allowed.  For the parton-distribution functions,
we have used the CTEQ6L1~\cite{CTEQ6PDFs} PDF set, and we have taken
$\sqrt{s} = 7$~TeV.  The cross-section enhancement factors are
much larger for fermions than for scalars, as one would expect, and
each decreases slowly with increasing $m_X$.
For $m_X \sim 375$~GeV, which corresponds to the value of $M_{jjj}$ for
which the greatest excess was observed by the CMS collaboration, the
cross-section for the pair production of a scalar $\rb{15}$ is roughly
twice that for a gluino.  For the fermionic $\rb{10}$ and scalar
$\rb{24}$, the corresponding enhancement factor is roughly ten, and for
the fermionic $\rb{15}'$ and scalar $\rb{35}$, it is far larger.

%-----------------------------------------------------------------------------
\section{Comparison to CMS Data\label{sec:CMSComparison}}
%-----------------------------------------------------------------------------

Having derived results for the pair-production cross-sections for an exotic
field $X$ in a higher representation of $SU(3)_c$,
we now assess the implications of the recent CMS  trijet resonance
search~\cite{CMSTrijetSearch} for such a particle.  In this search,
events with at least six jets were considered, and
invariant masses $M_{jjj}$ were reconstructed for all twenty possible
combinations of three jets from among the six highest-$p_T$ jets in each such
event.  A series of event-selection criteria were then imposed, including a
cut on $M_{jjj}$ designed to reduce the combinatoric background.
The prediction for the SM background, to which the QCD background
provides the dominant contribution, was obtained by fitting an exponential
function of the form $\exp(P_0 + P_1 M_{jjj})$ to the $M_{jjj}$ distribution
obtained for $N_j = 4$ events in experimental data (where $N_j$ denotes the
number of jets in the event), and then subsequently rescaling the normalization
coefficient $P_0$ on the basis of the average scalar $p_T$ of the triplets
observed in the $N_j \geq 6$ data.
The prediction for the signal, which was assumed to be from a decaying
gluino of mass $M_{\widetilde{g}}$ in a supersymmetric model with R-parity
violation, was obtained by simulating event samples for a broad range of
$M_{\widetilde{g}}$.  For each value of $M_{\widetilde{g}}$, an acceptance
$k(M_{\widetilde{g}})$, representing the effect of the event-selection criteria
on the signal sample, was obtained.  From the form of the acceptance function
$k(M_{\widetilde{g}})$, which was found to be approximately quadratic in
$M_{\widetilde{g}}$, limits on the gluino production cross-section in such
theories --- and therefore a limit on $M_{\widetilde{g}}$ --- was derived.

In order to estimate the corresponding exclusion limits on the mass
of a particle $X$ in a given representation of $SU(3)_c$, we make the
assumption that the next-to-leading-order (NLO) K-factor for $X$ pair
production at the $\sqrt{s} = 7$~TeV LHC
is essentially the same as the K-factor for gluino pair production
for any given value of $m_X$ within the range of interest.  Under this
assumption, we may obtain the NLO cross-section for the pair production of
$X$ by scaling the NLO cross-section for gluino production by the enhancement
factor displayed in Fig.~\ref{fig:EnhVsGluinoPlot}.  By comparing these
results to the observed 95\% CL limits on the production cross-section for
a heavy particle which decays primarily into three jets, we obtain our
exclusion limits on $m_X$ for each of the otherwise phenomenologically viable
$SU(3)_c$ representations for $X$ enumerated in Sect.~\ref{sec:RepTheory}.  
Note that we are also assuming that the cut acceptance is independent of the 
Lorentz and $SU(3)_c$ representation. 

\begin{figure}[t!]
\centerline{
\epsfxsize 3.25 truein \epsfbox {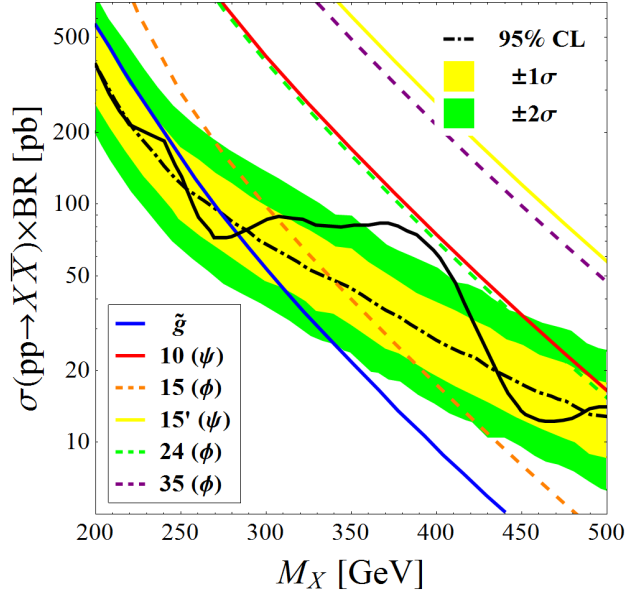} }
\caption{NLO cross-sections for the pair
production of an exotic field $X$ transforming under the
phenomenologically allowed representations of $SU(3)_c$ for which
$X$ decays exclusively to trijet final states, plotted as a
function of $m_X$.  Results are displayed for the cases in which $X$
is a fermionic $\rb{10}$ (solid red curve), a scalar $\rb{15}$ (dashed
orange curve), a fermionic $\rb{15}'$ (solid yellow curve), a
scalar $\rb{24}$ (dashed green curve), and a scalar $\rb{35}$
(dashed purple curve).
For reference, the corresponding NLO
cross-section for gluino production
in the limit of infinitely heavy squarks
(solid blue curve) has also been included for reference.
Also shown are the expected (dashed black curve)
and observed (solid black curve) 95\%~CL limits on the production
of a trijet resonance obtained in Ref.~\cite{CMSTrijetSearch} for
$\mathcal{L}_{\mathrm{int}} = 35\mathrm{~pb}^{-1}$
at the $\sqrt{s} = 7$~TeV LHC, along with the $\pm 1\sigma$
and $\pm 2\sigma$ bands on the expected limit.
\label{fig:GreenYellowPlot}}
\end{figure}

In Fig.~\ref{fig:GreenYellowPlot}, we display the NLO production cross-sections
for the production of a fermionic $\rb{10}$, a scalar
$\rb{15}$, a fermionic $\rb{15}'$, a scalar $\rb{24}$, and a scalar $\rb{35}$
over the range of $m_X$ pertinent to the CMS trijet analysis.  Also shown are
the expected and observed 95\% confidence-level (CL) exclusion limits on the
pair-production cross-section for a particle decaying essentially exclusively
to three jets obtained in Ref.~\cite{CMSTrijetSearch}, along with the $\pm 1\sigma$
and $\pm 2\sigma$ bands on the expected limit.  Any value of $m_X$ for which the
NLO cross-section exceeds the observed limit can be considered to be excluded at
95\%~CL.  From this figure, it is apparent that a
scalar $\rb{15}$ is clearly excluded for $m_X \lesssim 310$~GeV.
For the case in which $X$ is a fermionic $\rb{15}'$ or a scalar $\rb{35}$,
the exclusion limit on
$m_X$ from CMS data extends slightly beyond the mass range for which the
exclusion contour is displayed in Ref.~\cite{CMSTrijetSearch}.
One can estimate the exclusion limit on a fermionic $\rb{15}'$ or scalar
$\rb{35}$ by examining
where the extrapolated curve corresponding to the {\it expected} 95\%~CL limit and
the NLO pair-production cross-section intersect.  Based on this prescription, we
find that the CMS data exclude a fermionic $\rb{15}'$ with a mass
$m_X \lesssim 680$~GeV and a scalar $\rb{35}$ with a mass $m_X \lesssim 660$~GeV.
Similarly, the data can be interpreted as excluding a fermionic $\rb{10}$ and
scalar $\rb{24}$ with $m_X \lesssim 530$~GeV and $m_X \lesssim 520~$~GeV, respectively.
However, it is important to note that for $m_X \sim 375-400$~GeV, which corresponds to
the range of $M_{jjj}$ for which the CMS collaboration reported its greatest excess, the
estimated production cross-section for a fermionic $\rb{10}$ or scalar $\rb{24}$
only marginally exceeds the observed $95\%$~CL limit.   Given the uncertainties in
the NLO estimate for $X$ production, \etc, a fermionic $\rb{10}$ or scalar $\rb{24}$
with a mass $m_X \sim 375-400$~GeV can therefore also be interpreted as being
consistent with the data.

Even more intriguing, however, is the fact that
the CMS collaboration {\it did} report a $1.9\sigma$ excess in the number
of observed jet triplets in the invariant-mass range
$350\mathrm{~GeV} \lesssim M_{jjj} \lesssim 450\mathrm{~GeV}$.  Specifically,
an excess of approximately 30 total jet triplets over an expected SM background of
approximately 120 jet triplets was observed in this range.
We find that the distribution of these excess events as a function of $M_{jjj}$ can be
reasonably well modeled by a Gaussian centered around $M_{jjj} \sim 380$~GeV, with
a width of approximately 15~GeV.  Given the results for the acceptance function
$k(M_{\widetilde{g}})$ obtained by the CMS collaboration, we find that this
excess is roughly $\sim 2.5$ times larger than that which would be expected for a
gluino with $M_{\widetilde{g}} \approx 380$~GeV.  In other words, the observed excess
could be accounted for by a particle with similar production and decay phenomenology
to that of a gluino in an R-parity-violating supersymmetry scenario, but with a
production cross-section approximately $2.5$ times larger than that for such a particle.

We observe that while it is therefore improbable (though perhaps still
possible) that a gluino could account for the observed excess reported by CMS,
an additional field $X$ transforming in a higher representation of $SU(3)_c$ provides
a more reasonable fit to the data.
Indeed, it is apparent from the results shown in Fig.~\ref{fig:EnhVsGluinoPlot} that a
new scalar field $X$ with a mass $m_X \approx 380$~GeV which transforms in the $\rb{15}$
representation of $SU(3)_{c}$ would have just the right cross-section to account for
the observed excess.  However, we reiterate that due to the uncertainties in the
NLO K-factors, \etc, a Dirac fermion of comparable mass transforming
in the $\rb{10}$ representation or a scalar transforming in the $\rb{24}$
representation could potentially also explain the observed excess.

%-----------------------------------------------------------------------------
\section{Conclusions\label{sec:Conclusions}}
%-----------------------------------------------------------------------------

In this paper, we have shown that the multi-jet-resonance study performed by the
CMS collaboration, which was motivated primarily as a search for a light gluino in
R-parity-violating supersymmetric models, can also be used to probe a variety of
other scenarios for new physics.  In particular, we have shown that there exist
several representations of $SU(3)_c$ for which a heavy exotic field $X$ transforming
under one of these representations is likewise forced by gauge and Lorentz invariance
alone to decay essentially exclusively to a trijet final state.  We have examined
the detection prospects for such a particle, and have used the results of the CMS
study to derive exclusion limits on the representations of $SU(3)_c$ under which
$X$ could feasibly transform, given additional constraints from renormalization-group
running, \etc~

Furthermore, we have shown that the $\sim 2\sigma$ excess at
a trijet invariant mass of $M_{jjj} \approx 375$~GeV reported in that study can be
explained by the presence of an scalar transforming as a $\rb{15}$ or $\rb{24}$
of $SU(3)_c$, or by a Dirac fermion transforming as a $\rb{10}$.  (The scalar $\rb{15}$
provides the best fit to the data.)  By contrast, the production cross-section for an
R-parity-violating gluino is substantially smaller, and such a particle therefore
offers a less compelling explanation for the observed excess.
As further data is accumulated by the ATLAS and CMS experiments, it will be
interesting to see whether that data corroborate this potential signal of new physics,
and if so, whether they remain consistent with the interpretation we have suggested here.
More generally, any future excess or peak observed in a multi-jet invariant-mass
distribution (in events with or without the presence of substantial missing energy)
is amenable to an analysis of the sort.

It should be noted that the assumptions we have made in Sect.~\ref{sec:CMSComparison}
concerning the K-factors for the production cross-sections for fields
in higher representations of $SU(3)$ are certainly reasonable in the
absence of explicit NLO calculations, and to date, such calculations have
yet to be performed.  We note, however, that the true NLO K-factors for
these cross-sections may differ --- perhaps significantly --- from those
adopted in this study.  Indeed, examples of situations in which the
result of a full NLO calculation turned out to differ significantly from
the projected result adopted for the purpose of preliminary analysis do
exist in the literature~\cite{GaoEtAlNLO}.  Given this, the results
displayed in this study can be taken as sufficient motivation for
detailed NLO analyses of the pair-production cross-sections for fields in higher
representations of SU(3).  Indeed, the results of such analyses may prove
crucial for distinguishing between new-physics explanations for a given
signal or data anomaly observed at the LHC.

If any multi-jet resonance is indeed confirmed at the LHC, the next step
would be to identify complementary channels in which one could obtain evidence
that such a resonance is indeed due to a particle transforming under one of the
representations of $SU(3)_c$ which intrinsically forbid all decays to
anything other than a trijet final state.
Fortunately, in any grand unified (or even partially unified) theory,
any field transforming under a higher representation of $SU(3)_c$ would
necessarily have to be incorporated into some representation of the unified group.
Therefore, if some such field is truly responsible for a given
multi-jet resonance, each different unification scenario which could accommodate
that field would generically provide a prediction for other new particles which could
also potentially be discovered at the LHC.  These predictions become particularly
explicit for fields in representations which would dramatically alter the running of
$\alpha_s$ to the extent that new physics would be required only slightly above the
TeV scale to regulate divergences in the theory.
Even in less extreme situations,
however, any effect on the running of $\alpha_s$ could potentially alter the
unification scale, and signals of particles transforming in higher representations of
$SU(3)_c$ could thus provide valuable insight into the nature of our universe at high
scales.

\section{Acknowledgments}

We are grateful to J.~Bramante and D.~Yaylali for useful discussions.  JK and BT are
supported in part by the Department of Energy under Grant~DE-FG02-04ER41291.
The work of AR is supported in part by NSF grant PHY-0653656. AR would also like
to thank the Aspen Center of Physics, where part of this work was completed, for
hospitality.

\end{document}